\documentclass[onecolumn]{IEEEtran}
\usepackage{amsmath}
\usepackage{theorem}
\usepackage{amsmath}
\usepackage{mathrsfs}
\usepackage{accents}
\usepackage{url}
\usepackage{algorithm}
\usepackage{algorithmic}
\usepackage{epsfig}
\usepackage{amsfonts}
\usepackage{amssymb}
\usepackage{mathrsfs}
\usepackage{euscript}
\usepackage{graphicx}
\usepackage{multirow}
\usepackage{cite}
\usepackage{epsfig}
\usepackage{graphics}
\usepackage[protrusion=true,expansion=true]{microtype} % Better typography
\usepackage{graphicx} % Required for including pictures
\usepackage{wrapfig} % Allows in-line images
\usepackage{epstopdf}
\usepackage{amsmath}
\usepackage{array}
\usepackage{booktabs}
\usepackage{tabularx}
\usepackage{tabulary}

\newtheorem{lemma}{Lemma}

\usepackage{algorithm}
\usepackage{algorithmic}
\usepackage{setspace}
\usepackage[nodayofweek,level]{datetime}
\newcommand{\mydate}{\formatdate{2}{07}{2018}}
\newcommand\blfootnote[1]{%
	\begingroup
	\renewcommand\thefootnote{}\footnote{#1}%
	\endgroup
}
\begin{document}
	
	\begin{titlepage}

		\begin{tabular}{l r}
			
			\includegraphics[scale=0.3]{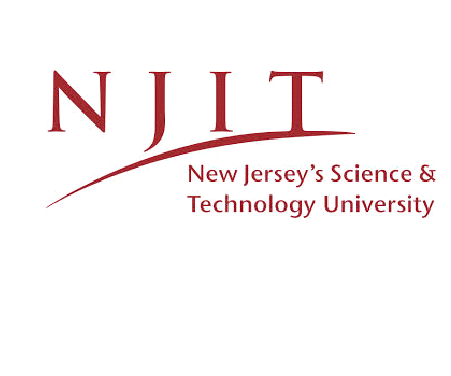} \hspace{10cm} & \includegraphics[scale=0.3]{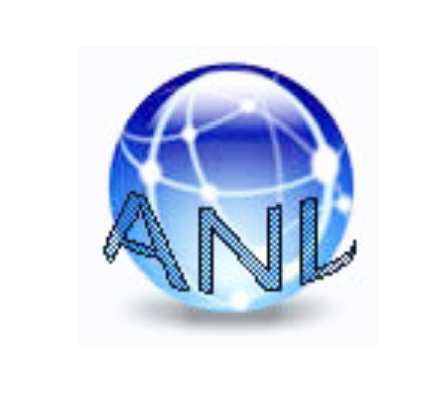}
			
		\end{tabular}
		
		\begin{center}
			\vspace{10mm}
			
			\textsc{\LARGE Energy Efficient Resource Allocation in EH-enabled CR Networks for IoT}\\[4cm]
			
			{\Large \textsc{Ali Shahini}}\\ % Author name
			{\Large \textsc{Abbas Kiani}}\\
			{\Large \textsc{Nirwan Ansari}}\\ % Author name
			[4cm]
			
			{}
			{\textsc{TR-ANL-2018-001}\\
				%\large \today}\\[3cm]
				\large \mydate} \\[4cm]
			
			{\textsc{Advanced Networking Laboratory}}\\
			{\textsc{Department of Electrical and Computer Engineering}}\\
			{\textsc{New Jersey Institute of Technology}}\\[1.5cm]
			\vfill
			
		\end{center}
		
	\end{titlepage}
\onecolumn

\begin{abstract}
With the rapid growth of Internet of Things (IoT) devices, the next generation mobile networks demand for more operating frequency bands. By leveraging the underutilized radio spectrum, the cognitive radio (CR) technology is considered as a promising solution for spectrum scarcity problem of IoT applications. In parallel with the development of CR techniques, Wireless Energy Harvesting (WEH) is considered as one of the emerging technologies to eliminate the need of recharging or replacing the batteries for IoT and CR networks. To this end, we propose to utilize WEH for CR networks in which the CR devices are not only capable of sensing the available radio frequencies in a collaborative manner but also harvesting the wireless energy transferred by an Access Point (AP). More importantly, we design an optimization framework that captures a fundamental tradeoff between energy efficiency (EE) and spectral efficiency (SE) of the network. In particular, we formulate a Mixed Integer Nonlinear Programming (MINLP) problem that maximizes EE while taking into consideration of users' buffer occupancy, data rate fairness, energy causality constraints and interference constraints. We further prove that the proposed optimization framework is an NP-Hard problem. Thus, we propose a low complex heuristic algorithm, called INSTANT, to solve the resource allocation and energy harvesting optimization problem. The proposed algorithm is shown to be capable of achieving near optimal solution with high accuracy while having polynomial complexity. The efficiency of our proposal is validated through well designed simulations.
\end{abstract}

\blfootnote{The authors are with Advanced Networking Laboratory, the Helen and John C. Hartmann Department of Electrical and Computer Engineering, New Jersey Institute of Technology, Newark, NJ 07102.\protect~E-mail: {ali.shahini, abbas.kiani, nirwan.ansari}@njit.edu. This work was supported in part by NSF under grant no. CNS-1320468.}

\section{Introduction}\label{sec:Introduction}
\IEEEPARstart{I}{nternet} of Things (IoT) is an emerging computing concept in which everyday physical objects such as sensors, home appliances, wearable electronics, and  environmental monitors are connected to the Internet without external human intervention. Owing to the wide variety of fields for IoT applications including smart houses, connected cars, smart cities, wearables, smart retails, and connected health, the number of connected devices has increased tremendously and anticipated to be more than 50 billions by 2020~\cite{0036,XilongIoT,MagazinePPT2}. Although the traditional  energy-constrained wireless networks are powered by fixed energy sources like batteries, it may be expensive and inconvenient to replace and recharge batteries as the number of IoT devices increases. Therefore, one of the dominant barriers to implementing IoT is supplying adequate energy to operate the network in a self-sufficient manner \cite{HuangAnsari}. Wireless Energy Harvesting (WEH), as one of the most promising solutions, is proven to eliminate the need of recharging or replacing the batteries entirely. Unlike other types of green energy sources (e.g., wind, solar, and vibrations), WEH does not depend on nature, and is thus a reliable source of energy for IoT devices \cite{GreenBook}. The WEH schemes are classified into three categories: energy harvesting from unknown source, anticipated source and intended Wireless Energy Transmission (WET), respectively. While the two formers are not efficient because the amount of ambient wireless energy in the environment is generally low and inconsistent, the latter, which can utilize the power transmitters (e.g., Powercast TX91501) \cite{Powercast}, is much more efficient.

The rapid growth of higher data rate devices and wireless applications is likely to demand for more operating frequency bands. Although the frequency spectrum is currently scarce, a considerable amount of the radio spectrum is greatly underutilized \cite{SurveySensing,00121}. Thus, the dynamic spectrum access capabilities of Cognitive Radio (CR) can be considered as a potential solution to address the aforementioned challenge by utilizing the spectrum holes, i.e., underutilized spectrum band. In fact, the wireless devices are equipped with cognitive capability to search for the spectrum holes and try to utilize them opportunistically in order to transmit data without causing interference to licensed users. Therefore, spectrum sensing, which is the process of detecting the spectrum holes, plays a crucial role. Owing to the effects of fading and shadowing, the performance of single spectrum sensing (local spectrum sensing) is generally unreliable. In this regard, the Cooperative Spectrum Sensing (CSS) method is applied to improve the performance of sensing by combining the observations of spatially located users. Current communication technologies cannot provision the future growth of numerous IoT devices. Therefore, transforming the IoT network into a cognitive based IoT network is essential to utilizing the available spectrum opportunistically \cite{0045}.
%\textbf{\cite{0032,0033,0035,0037,0038}}

\subsection{Related Works}

The increase in energy consumption of IoT devices is much faster than the development of battery technology. Thus, energy efficiency (EE) of IoT networks has emerged as a major research issue and becomes a trend for the future design of wireless communications. EE has been brought up and discussed in IoT networks \cite{0044,0034,0039,0047}. In particular, an energy efficient architecture has been proposed in \cite{0044} for IoT networks where the sensors' sleep intervals are predicted based on their remaining energy. Sharma \textit{et al.}~\cite{0034} presented an energy efficient approach for device discovery in 5G-based IoT using multiple Unmanned Aerial Vehicles (UAVs). Zhang \textit{et al.} \cite{0039} proposed an integrated structure to enhance the EE of IoT networks. They optimized the EE of the whole system by considering the wireless and wired parts at the same time. Alnakhli \textit{et al.} \cite{0047} proposed a mechanism to jointly maximize the spectrum and energy efficiency for device to device communications enabled wireless networks.

IoT and CR networks are evolving technologies and the CR utilization in IoT is becoming an important issue. However, few works have discussed the CR capabilities (like cooperative spectrum sensing) for IoT networks. State of the arts on cognitive Machine to Machine (M2M) communications from a protocol stack perspective has been reviewed in~\cite{0046}. Majumdar \textit{et al.} \cite{0048} also proposed a packet size optimization mechanism for cognitive radio based IoT networks where they consider the tradeoff that exists in terms of EE and overhead delay for a given data packet length. Throughput maximization has been proposed in \cite{Sarnoffam} for energy harvesting enabled CR networks. Moreover, Hu \textit{et al.} \cite{0049} proposed a cognitive code division multiple access scenario by combining the concept of CR with dynamic spectrum bands and CDMA for IoT networks.

Wireless energy harvesting and transfer technologies have recently emerged as a practical and effective solution to address the issues of energy efficiency. Some recent works \cite{004,006,009,Elsevieram,003} have brought up this issue. Li \textit{et al.}~\cite{004} proposed a framework where network coding is applied to an IoT network to reduce IoT energy consumption. Kawabata \textit{et al.} \cite{006} considered a relay selection problem for energy harvesting and proposed a new scheme for energy harvesting relay selection which is based on the residual energy at each relay's battery. Song \textit{et al.} \cite{009} studied a tradeoff between Quality of Service (QoS) provisioning and the energy efficiency for IoT networks. Joint spectrum allocation and energy harvesting optimization has been proposed in \cite{Elsevieram} for green powered heterogeneous cognitive radio networks. Moreover, Liu \textit{et al.} \cite{003} proposed a wireless energy harvesting protocol for an underlay cognitive relay in which the secondary users are assumed to harvest energy from the primary network. 

\subsection{Contributions}
None of the existing works consider a model based on the advantages of both WEH and CSS in the context of CR based IoT. Therefore, this study aims to address the aforementioned issue by proposing a system model that not only leverages WEH and CSS, but also is designed to maximize the EE for the CR based IoT. In particular, we optimize the tradeoff between the EE and spectral efficiency (SE), thereby taking into consideration of the limits of spectrum resources. The main contributions of this paper are summarized as follows.
 
\begin{itemize}
	\item We propose a CR based system model for IoT using wireless energy harvesting and cooperative spectrum sensing to tackle two vital challenges of IoT networks, i.e., supplying adequate energy to operate the network in a self-sufficient manner, and providing enough radio spectrum to accommodate the massive growth of devices. To this end, we consider a time switching model in which the devices participate in the CSS process at the beginning of each time slot and report their results to an Access Point (AP) to identify idle frequency bands. Then, they harvest energy that is intentionally transmitted by the AP and finally transmit their data using the harvested energy. Since the users operate in a time switching fashion, one should notice that there is a tradeoff between the length of the energy harvesting process and data transmission part. Therefore, we focus on optimizing the EE and SE tradeoff of the network by optimizing the length of energy harvesting process in each time slot while ensuring data rate requirements of the devices and the fairness in channel allocation among the users.
	\item We formulate an MINLP problem to maximize the tradeoff between the EE and SE while taking into consideration of practical limitations such as data rate fairness, energy causality constraints, interference constraints, stochastic energy harvesting rates and imperfect spectrum sensing. It is proved that the optimization framework is an NP-Hard problem.
	\item We thus propose a low complex heuristic algorithm, referred to as  joINt Sub-channel  allocaTion And eNergy harvesting opTimization (INSTANT), to solve the sub-channel allocation and energy harvesting optimization problem. The proposed algorithm is shown to be capable of achieving near optimal solution with high accuracy while having polynomial complexity. 
\end{itemize}
The rest of the paper is organized as follows. We propose the system model in Section II. Section III describes the problem formulation. The energy efficiency maximization problem is presented in Section IV. We propose our heuristic algorithm in Section V. Finally, simulation results and conclusion of this paper are presented in Sections VI and VII, respectively.

\section{System Model}\label{sec:System Model}
\begin{figure} [t]
	\centering
	\includegraphics[width=9cm,height=5cm]{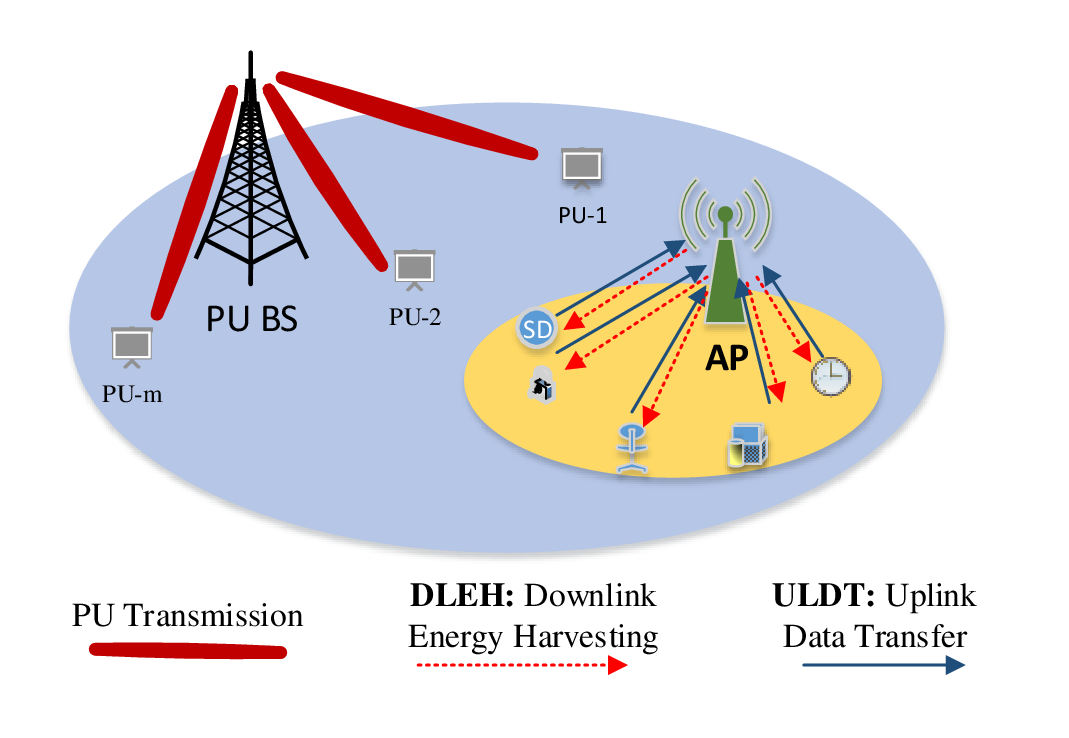}
	\caption{The system model.}
	\label{fig:SysModel}
\end{figure}
Consider two cellular systems, one of which is for $M$ primary users (PUs) denoted by ${\cal M} = \{ 1,2,...,M\}$ and the other is for a time-slotted CR-based IoT network comprising $K$ self-powered devices represented by ${\cal K} = \{ 1,2,...,K\}$ (Fig. (\ref{fig:SysModel})). In this model, the devices opportunistically utilize the licensed radio spectrum of the PUs via an AP. The AP is equipped with one Fusion Center (FC) to centrally process the sensing results of the users and one energy transmitter to broadcast energy signals to its associated devices. Each device does local spectrum sensing by a low complex Energy Detector (ED) concerning the presence of PUs. We assume that the spectrum sensing results of users are independent \cite{Shroff} and each user is permitted to sense any number of sub-channels. The local spectrum sensing results are sent to the FC located in the AP. Then, the FC applies a CSS strategy (to reduce the sensing errors) to achieve final decisions regarding availability of licensed sub-channels. Denote ${\cal N} = \{ 1,2,...,N\}$ as the subset of available sub-channels identified by the CR based IoT network for data transmission.

The devices are also considered to be self-powered in which they are equipped with wireless energy harvester devices (e.g., P2110B Powercast receiver \cite{PowercastReceiver}) to exclusively harvest energy from wireless energy signals intentionally transmitted by the AP. In fact, the AP broadcasts a deterministic energy signal to power the nearby users over the downlink channel and then receives data from these users transmitted via the uplink channel. The users store their energy in their temporary energy storage devices (e.g., capacitors). However, the energy harvesting process and energy consuming process cannot be done simultaneously in such devices. Thus, due to the \textit{energy half-duplex constraint}, while the user transmits data via the uplink channel, its energy harvester pauses \cite{00146}.

\begin{figure} [t]
	\centering
	\includegraphics[width=9cm,height=6cm]{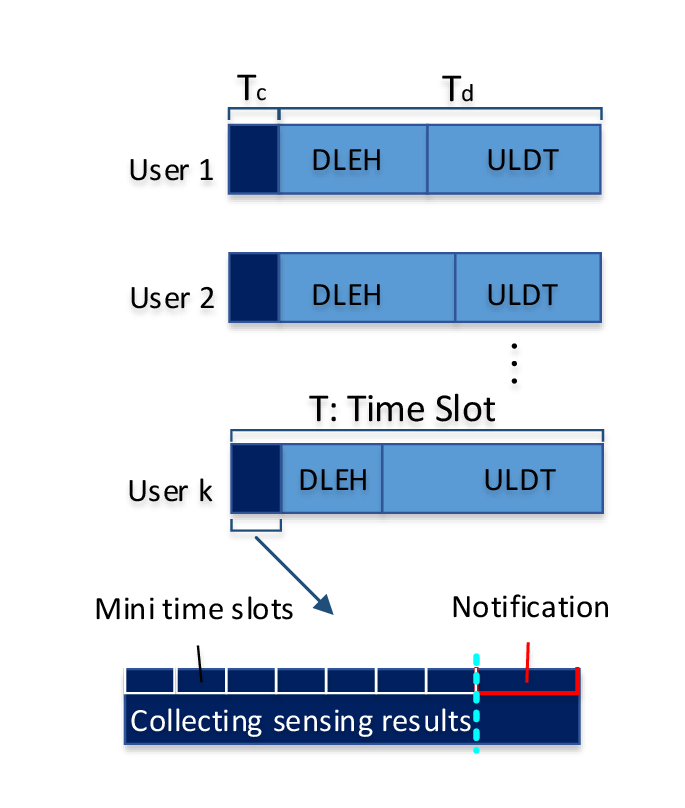}
	\caption{The time slot model.}
	\label{fig:TimeSlot}
\end{figure}

Basically, each time slot with duration $T$ is partitioned into a control slot $T_c$ and a data slot $T_d$ (Fig. (\ref{fig:TimeSlot})). The length of the control slot is called the sensing overhead and is constant for all devices \cite{Shroff}. The fixed control slot period is devoted for CSS and reporting the optimization results from the AP in each slot. We assume that at the beginning of the $t^{th}$ time slot, the device $k \in \mathcal{K}$ has residual energy $E_{k,t}^{res}$ that is enough for spectrum sensing during the control slot. Meanwhile, the data time slot is divided into two non-overlapping parts, namely, Downlink Energy Harvesting (DLEH) and Uplink Data Transmission (ULDT). We define the harvesting ratio for the $k^{th}$ device as $\mu_{k}$, $\forall k \in \mathcal{K}$, which determines the fraction of data slot devoted to the energy harvesting process. Nevertheless, in the slotted operating mode, with more time spent on DLEH, less time remains for ULDT, thus degrading the achievable throughput. Hence, there exists a tradeoff between DLEH and ULDT durations. Meanwhile, the users are assumed to have perfect knowledge of Channel State Information (CSI) between their transmitters and the AP receiver.

\section{Problem Formulation}
\subsection{Energy Consumption} 
Denote $\rho_k,~\forall k \in \mathcal{K}$, as the energy harvesting rate of the $k^{th}$ user. We assume that energy harvesting rates are independent random variables with a common general distribution, where the average energy harvesting rate of the $k^{th}$ user is $E[ {{\rho _k}}]\triangleq{{\rho }_k^{av}},~\forall k \in \mathcal{K}$. Thus, the amount of energy harvested by the $k^{th}$ device in the $t^{th}$ time slot is $E_{k,t}^{har} = {\rho }_k^{av} \mu_k T_d$, where $\mu_k T_d$ is the amount of time in the time slot devoted for energy harvesting. Since we assume that the users have enough residual energy at the beginning of each time slot to operate spectrum sensing, the following inequality must hold
\begin{equation}
E_{k,t}^{res} - E_{k,t}^{sen} + E_{k,t}^{har} - E_{k,t}^{tr} \geqslant E_{k,t+1}^{sen},
\end{equation}
where $E_{k,t}^{sen}$ denotes energy expenditure of the $k^{th}$ user to sense the spectrum in time slot $t$. $E_{k,t}^{tr}$ is the amount of energy consumed by the $k^{th}$ user to transmit data in the $t^{th}$ time slot. However, the energy of sensing does not change from time slot to time slot for each user, $E_{k,t}^{sen} = E_{k,t+ 1}^{sen}$. Thus, the energy causality constraint for the network of $K$ users is given by
\begin{equation} \label{energycausality}
E_{k,t}^{res} + E_{k,t}^{har} \geqslant E_{k,t}^{tr} + 2E_k^{sen},\forall k \in \mathcal{K}.
\end{equation}
Accounting for the harvested and the consumed energies, the residual energy of the $k^{th}$ user $\forall k \in \mathcal{K}$ at the beginning of the next time slot is updated as follows
\begin{equation}
E_{k,t + 1}^{res} = E_{k,t}^{res} + E_{k,t}^{har} - E_{k,t}^{tr} - E_k^{sen}.
\end{equation}
To calculate the total energy consumption of one user in one time slot, one should consider the energy of transmission, sensing energy and the consumed energy for being idle. Based on the energy causality constraint in Eq.~(\ref{energycausality}), the maximum transmission energy of the $k^{th}$ user in time slot $t$ is given by
\begin{equation}
E_{k,t}^{tr} = E_{k,t}^{res} + E_{k,t}^{har} - 2E_{k,t}^{sen}.
\end{equation}
Denote $E_{k,t}^{idle}$ as the energy of the idle mode for the $k^{th}$ user in time slot $t$. Therefore, the energy consumption of the $k^{th}$ user in the $t^{th}$ time slot depends on the energy of transmission, harvested energy, spectrum sensing energy consumption and energy for remaining idle, i.e.,
\begin{equation}
E_{k,t}^{con} = \underbrace {E_{k,t}^{res} + {\rho }_k^{av}{\mu_k}{T_d} - 2E_{k,t}^{sen}}_{E_{k,t}^{tr}} + E_{k,t}^{sen} + \underbrace {{P_{k,t}^{idle}}({T_c} - {\tau _{{s_k}}})}_{E_{k,t}^{idle}}.
\end{equation}
Thus, the total energy consumption of $K$ users in each time slot as a function of the harvesting ratio can be written as
\begin{equation} \label{ETotal}
E_{total}^{con}({\mu _k}) = \sum\limits_{k \in \mathcal{K}} {\left( {{\rho }_k^{av}{\mu_k}{T_d} + E_{k}^{res} - E_{k}^{sen} + E_{k}^{idle}} \right)}.
\end{equation}

\subsection{Achievable Throughput}\label{throughputsubsec}

To measure the Shannon capacity for each user in the system, we define ${t_k^{tr}}({\mu _k}) \triangleq (1-\mu_k)T_d$ and $E_k^{tr}({\mu _k})\triangleq{E_k^{res} + {\rho }_k^{av}`{\mu_k}{T_d} - 2E_k^{sen}}$ as the transmission time and the transmission energy of the $k^{th}$ user as a function of the harvesting ratio $\mu_k$. Thus, the transmission rate of the $k^{th}$ user over the $n^{th}$ sub-channel can be written as 
\begin{equation}
{r_{k,n}({\mu _k})} = \frac{{t^{tr}}({\mu _k})}{T}lo{g_2}\left( {1 + \frac{{\left| {{h_{k,n}}} \right|}^2E_k^{tr}({\mu _k})}{{\Gamma (B{N_0} + I_k)}{t_k^{tr}}({\mu _k})}} \right),
\end{equation}
where ${{h_{k,n}}}$ denotes the channel gain of the $k^{th}$ user over the $n^{th}$ sub-channel, $B$ is the  bandwidth  of  each  OFDM  sub-channel, $N_0$ represents the additive white Gaussian noise, $I_k$ is a measured interference introduced to the $k^{th}$ user caused by the PUs' signals, and $\Gamma$ denotes the SNR gap associated with the bit-error-rate (BER) of un-coded MQAM.

Denote $g_{k,n}$ as the binary variable to allocate of the $n^{th}$ sub-channel to the $k^{th}$ user. In other words, $g_{k,n} = 1$ if there is an allocation, and $g_{k,n} = 0$ otherwise. Moreover, for the sake of convenience, we define ${H_{k,n}} \triangleq \frac{{{{\left| {{h_{k,n}}} \right|}^2}}}{{\Gamma (B{N_0} + {I_k})}}$. Therefore, the total transmission rate of the $k^{th}$ user over all available sub-channels, $N$, is given by
\begin{equation}  \label{Rkkk}
{R_k({\mu _k},g_{k,n})} = \sum\limits_{n = 1}^N {{g_{k,n}}\frac{{t_k^{tr}}({\mu _k})}{T}lo{g_2}\left( {1 + {H_{k,n}}\frac{{E_k^{tr}({\mu _k})}}{{t_k^{tr}}({\mu _k})}} \right)}.
\end{equation}
%Then, the effective rate can be written as
%\begin{equation}
%{C_{k}} = min({Q_k},\overline {R_k({\mu _k},g_{k,n})}{t_k^{tr}}({\mu _k})),
%\end{equation}
%where $Q_k$ is the $k^{th}$ IoT buffer size and is considered to be sufficiently large. Thus, the total number of transmitted bits of the $k^{th}$ IoT user is ${\overline{R_k({\mu _k},g_{k,n})}{t_k^{tr}}({\mu _k})}$.

%Since the energy harvesting rate is stochastic, the achievable throughput in each time slot becomes a random variable. Thus
%\begin{equation}
%{\rm E}[r_{k,n}({\mu _k})] = \frac{{t^{tr}}({\mu _k})}{T}{\rm E}\left[ {lo{g_2}\left( {1 + \frac{{\left| {{h_{k,n}}} \right|}^2 {E_k^{tr}({\mu _k})}}{{\Gamma (B{N_0} + I_k)}{t_k^{tr}}({\mu _k})}} \right)} \right].
%\end{equation}

%\begin{equation} \label{RTotal}
%\overline{R_{total}({\mu _k},g_{k,n})} = {\sum\limits_{k \in \mathcal{K}} {\overline{R_k({\mu _k},g_{k,n})}{t_k^{tr}}({\mu _k})} }.
%\end{equation}
\section{Energy Efficiency Maximization}
In this section, we formulate an optimization problem to maximize the energy efficiency of the users while taking into consideration of their buffer occupancy. 
The energy efficiency is defined as the ratio of the total transmission rate to the total energy consumption, and is measured in unit of bits/sec/Joule. Recalling the total throughput in Eq.~(\ref{Rkkk}) and the total energy consumption in Eq.~(\ref{ETotal}), the energy efficiency of the $k^{th}$ user is
\begin{equation}
{{E_{eff_k}}({\mu _k},{g_{k,n}})}  =  {\frac{{{{R_k}({\mu _k},{g_{k,n}})}}}{{\rho _k^{av}{\mu _k}{T_d} + E_k^{res} - E_k^{sen} + E_k^{idle}}}}.
\end{equation}

Meanwhile, it is possible to allocate spectral resources to the users which do not have enough data in their buffers to transfer, thus resulting in a waste of spectrum resources. Let random variable $X_k$ represent the number of bits in the buffer of the $k^{th}$ user. We also assume that the random variables $X_k,~\forall k \in \mathcal{K}$, are independent and have a common general distribution with the average of $E({X_k}) \triangleq \overline {{X_k}}$.
In order to efficiently utilize the spectrum, the users should not receive spectral resources more than their data in the corresponding buffers. To this end, we incorporate the probability $P({X_k} \geqslant  {{R_k}({\mu _k},{g_{k,n}})} t_k^{tr}({\mu _k})),~\forall k \in \mathcal{K}$, into the objective function which ensures efficient utilization of the spectrum (spectral efficiency). Let  ${{S_{ef{f_k}}}}  \triangleq {\eta _k}P({X_k} \geq {{R_k}({\mu _k},{g_{k,n}})} t_k^{tr}({\mu _k}))$, where ${\eta _k}$ is the weight of the spectral efficiency in the objective function. 
However, maximum EE and SE cannot be obtained simultaneously; in fact, there is a trade-off between EE and SE. 
Therefore, we propose an optimization framework that optimizes the tradeoff between EE and SE in the CR-based IoT network with downlink energy harvesting, and is formulated as
\begin{equation} \label{GeneralOpt}
\begin{array}{*{20}{l}}
{\mathop {\max }\limits_{{\mu _k},{g_{k,n}}} }&{\sum\limits_{k \in \mathcal{K}} ({ {{E_{ef{f_k}}}({\mu _k},{g_{k,n}})}  + {{S_{ef{f_k}}}({\mu _k},{g_{k,n}})}} }) \\ 
{C1}&{{ {{R_k}({\mu _k},{g_{k,n}})} \ge h(E({X_k})),~\forall k \in \mathcal{K}} }\\ 
{C2}&{E_k^{res} + E_k^{har}\ge E_k^{tr} + 2E_k^{sen},~\forall k \in \mathcal{K}} \\ 
{C3}&{{T_d}\sum\nolimits_{k \in \mathcal{K}} {{\rho _k}{\mu _k}\le E_{MAX}^{AP},~\forall k \in \mathcal{K}} }\\ 
{C4}&{\sum\nolimits_{k \in \mathcal{K}} {P_k^{tr}({\mu _k}){I_{k,m}}} \le I_m^{th},~\forall m \in \mathcal{M}}\\ 
{C5}&{\sum\nolimits_{k \in \mathcal{K}} {{g_{k,n}} = 1,~\forall n \in \mathcal{N}} }\\ 
{C6}&{{g_{k,n}} \in \{ 0,1\} ,~\forall k \in \mathcal{K},~\forall n \in \mathcal{N}}&{} \\ 
{C7}&{E_k^{tr}\ge 0,~\forall k \in \mathcal{K}} \\ 
{C8}&{0 < {\mu _k} < 1,~\forall k \in \mathcal{K}} 
\end{array}
\end{equation}
C1 ensures minimum transmission rates, i.e.,  $h(E({X_k}))$, for all the users where $h$ is an increasing function for given $E({X_k})$.
In fact, the minimum data rate requirement of each user is defined as an increasing function of the average number of bits in its buffer. By doing so, we not only satisfy a minimum data rate for all the users but also consider a fair resource allocation based on the buffer occupancies.  
C2 means that the energy causality constraint should be held. C3 imposes that the total energy harvesting of users are less than the maximum transmitted energy of the AP. C5 and C6 imply that each sub-channel is allocated to no more than one device. C7 specifies that the transmission energy of each user should not have a negative value. C8 imposes the harvesting ratio to be a fraction of one time slot. Meanwhile, C4 specifies that the total interference to the $m^{th}$ PU must be less than a given threshold where $P_k^{tr}({\mu _k})$, the transmission power of the $k^{th}$ user, can be written as $\frac{{E_k^{res} + E[{\rho _k}]{\mu _k}{T_d} - 2E_k^{sen}}}{{(1 - {\mu _k}){T_d}}}$. 
Moreover, the devices have CR capabilities and utilize the idle spectrum opportunistically via CSS. Therefore, spectrum sensing errors (e.g., mis-detection of the spectrum) can occur and lead to co-channel interference. The total interference introduced to the $m^{th}$ PU by the $k^{th}$ user's transmission over the allocated sub-channels is given by \cite{00109}
\begin{equation}
{I_{k,m}} = \sum\limits_{n \in {\mathcal{N}_u}} {{P_{1,n}}I_{n,m}^k + } \sum\limits_{n \in {\mathcal{N}_a}} {{P_{2,n}}I_{n,m}^k},
\end{equation}
where ${\mathcal{N}_u}$ is the set of sensed unavailable sub-channels, ${\mathcal{N}_a}$ represents the set of sensed available sub-channels, ${P_{1,n}}$ denotes the probability that the CR based IoT network correctly identifies the $n^{th}$ unavailable sub-channel, and ${P_{2,n}}$ is the probability that the network makes a wrong decision regarding availability of the $n^{th}$ occupied sub-channel.

%$I_{k,m}$ is the total interference introduced to the $m^{th}$ PU by the $k^{th}$ IoT user's transmission. One part of this interference stems from imperfect spectrum sensing and the other part arises from the IoT transmission over different sub-channels. More details regarding $I_{k,m}$ can be found in \cite{Sarnoffam,00109}.
\begin{lemma} \label{[NP-Hard]}
	The general problem of EE maximization of the CR-based IoT network in Eq.~(\ref{GeneralOpt}) is an NP-hard problem when $\eta_{k}=0, ~ \forall k \in \mathcal{K}$.
\end{lemma}
\begin{IEEEproof}
	To prove the NP-hardness of the optimization problem, one can show that  the problem is reducible to one of the proven NP-hard problems. Let assume the decision variables $\mu_{k}, ~ \forall k \in \mathcal{K}$, are given and $\eta_{k}=0, ~ \forall k \in \mathcal{K}$. Then, the constraints C2, C3, C4, C7, and C8 are relaxed and the optimization problem in Eq.~(\ref{GeneralOpt}) is transformed to the following problem:
	
	\begin{equation} \label{NPHARD}
	\begin{array}{*{20}{l}}
	{\mathop {\max }\limits_{{g_{k,n}}} }&{\sum\limits_{k \in \mathcal{K}} {\frac{{ {{R_k}({\mu _k},{g_{k,n}})} }}{{\rho _k^{av}{\mu _k}{T_d} + E_k^{res} - E_k^{sen} + E_k^{idle}}}} } \\ 
	{s.t.}&{C1:\sum\nolimits_{k \in \mathcal{K}} {{g_{k,n}}{r_{k,n}} \geqslant h(E({X_k})),\forall k \in \mathcal{K}} }, \\ 
	{}&{C2:\sum\nolimits_{k \in \mathcal{K}} {{g_{k,n}} = 1,\forall n \in \mathcal{N}} } ,\\ 
	{}&{C3:{g_{k,n}} \in \{ 0,1\} ,\forall k \in \mathcal{K},\forall n \in \mathcal{N}} ,
	\end{array}
	\end{equation}
	In fact, the problem in Eq.~(\ref{NPHARD}) is a special case of the problem of Eq.~(\ref{GeneralOpt}) in which the second part of the objective function, i.e., $ {{S_{ef{f_k}}}}  \triangleq {\eta _k}P({X_k} \geq  {{R_k}({\mu _k},{g_{k,n}})} t_k^{tr}({\mu _k}))$, is removed. Thus, one can conclude that the reduced problem in Eq.~(\ref{NPHARD}) can be categorized as a Generalized Assignment Problem (GAP) which is a known NP-hard problem. Essentially, since the GAP problem is NP-hard, the optimization problem in Eq.~(\ref{GeneralOpt}) is also NP-hard.
\end{IEEEproof}

\section{Jo\textbf{IN}t \textbf{S}ub-channel Alloca\textbf{T}ion \textbf{A}nd E\textbf{N}ergy Harvesting Op\textbf{T}imization: \textbf{INSTANT}}

As discussed, finding an optimal solution to problem~(\ref{GeneralOpt}) is intractable. Therefore, to solve this problem, we propose a heuristic algorithm. The pseudo code for the proposed heuristic is shown in Algorithm~\ref{alg:1} where we follow two separated phases for optimizing the channel allocation as well as the harvesting ratio. Note that ${{S_{ef{f_k}}}}={\eta _k}\int_{R_k t_k^{tr}({\mu _k})}^{\infty}f_{X_k}(r)dr)$ where $f_{X_k}(r)$ is the probability distribution function of $X_k$.
Therefore, to solve problem~(\ref{GeneralOpt}), the distribution function of $X_k$ has to be known. For example, if $X_k$ is uniformly distributed between $a$ and $b$, $\int_{R_k t_k^{tr}({\mu _k})}^{\infty}f_{X_k}(r)dr=\frac{b-{R_k}({\mu _k},{g_{k,n}})t_k^{tr}({\mu _k})}{b-a}$. As another example, when $X_n$ follows an exponential distribution with parameter $\lambda$,
$\int_{R_k t_k^{tr}({\mu _k})}^{\infty}f_{X_k}(r)dr=e^{-\lambda{R_k}({\mu _k},{g_{k,n}})t_k^{tr}({\mu _k})}$.
In the proposed heuristic algorithm, we assume that the distribution function of $X_k$ is given.
Moreover, without loss of generality, we focus on the case that function $h(.)$ is given and linear.
\begin{algorithm} 
	\caption{Jo\textbf{IN}t \textbf{S}ub-channel Alloca\textbf{T}ion \textbf{A}nd E\textbf{N}ergy Harvesting Op\textbf{T}imization: \textbf{INSTANT}}		
	%Joint \textbf{S}ub-c\textbf{H}annel \textbf{A}llocation a\textbf{N}d Ener\textbf{G}y \textbf{HA}rvesting Opt\textbf{I}mization: \textbf{SHANGHAI}
	%	Joint \textbf{S}ub-channe\textbf{l} \textbf{a}llocatio\textbf{n} an\textbf{d} \textbf{E}nergy T\textbf{r}ansfer: SLANDER
	\label{alg:1}
	\begin{algorithmic}[1] 
		\STATE $\mu_k=\frac{\mu_k^{min}+1}{2}=~\forall k\in\mathcal{K}$
		\STATE $g_{k,n}=0~\forall k\in\mathcal{K}$ and $n\in\mathcal{N}$
		\STATE \textbf{Channel Allocation Phase}
		\STATE {$\mathcal{N}_0\gets\emptyset$}
		\STATE {$\mathcal{K}'\gets\mathcal{K}$}
		\FORALL {$n\in\mathcal{N}$}
		\STATE $\hat{k}=\underset {k\in\mathcal{K}'}{arg max}~R_k$
		\STATE $g_{\hat{k},n}=1$ and $\mathcal{N}_0\gets\mathcal{N}_0\cup n$
		\IF {$R_{\hat{k}}\geq h(E({X_{\hat{k}}}))$}
		\STATE {$\mathcal{K}'\gets\mathcal{K}'\setminus\hat{k}$}
		\ENDIF
		\ENDFOR
		\STATE {$\mathcal{N}\gets\mathcal{N}\setminus\mathcal{N}_0$}
		\STATE $\hat{o}=0$
		\FORALL {$n\in\mathcal{N}$}
		\FORALL {$k\in\mathcal{K}$}
		\IF {$o_k^n\geq\hat{o}$}
		\STATE $\hat{k}\gets k$
		\STATE $\hat{o}\gets o_k^n$
		\ENDIF
		\ENDFOR
		\STATE $g_{\hat{k},n}=1$
		\ENDFOR
		\STATE \textbf{Harvesting Ratio Optimization Phase}
		\STATE sort users in $\mathcal{K}''$ such that $\rho_1\leq\rho_2\leq...\leq\rho_K$
		\FORALL {$k\in\mathcal{K}''$}
		\STATE $\hat{\mu}_k\gets\underset {\mu_k^{min}\leq\mu_k\leq1}{arg max}~({\overline {E_{ef{f_k}}} +\overline{S_{ef{f_k}}}})$
		\IF {$\hat{R}_k\geq h(E({X_k}))$ and $\hat{E}_k\leq E_{MAX}^{AP}$ and $\hat{I}_k\leq I_m^{th}$}
		\STATE {$\mu_k=\hat{\mu}_k$}
		\ENDIF
		\ENDFOR
	\end{algorithmic} 
\end{algorithm}

As shown in algorithm~\ref{alg:1}, we initially set ${\mu _k} = \frac{{ \mu _k^{min} + 1}}{2}$ for all $k\in\mathcal{K}$, where $\mu_k^{min}=\frac{2E_k^{sen}-E_k^{res}}{E(\rho_k)T_d}$ is computed from constraint ${E_k^{tr}}\geq 0$.
Accordingly, having variables $\mu_k$ fixed to $\frac{\mu_k^{min}+1}{2}$, we follow our heuristic channel allocation phase.
Denote $R_k$ as the data rate of user $k$ and $\mathcal{N}_0$ as the set of the channels to be allocated to satisfy the minimum data rate requirement, i.e., $R_k\geq h(E(X_k))$. Let $\mathcal{K}'$ also be the set of users who cannot attain their minimum data rates yet.
We first allocate channels based on the channel conditions to satisfy constraint $R_k\geq h(E(X_k))$ for all the users (lines 6-12).
In fact, among the users with unsatisfied minimum data rate, a channel is allocated to the user that has the maximum rate on that channel.
Then, for the remaining channels, we search over all the users to find a favorite user to allocate each channel (lines 15-23). As shown in the algorithm, the favorite user $\hat{k}$ for channel $n$ is one that achieves the maximum increase in the objective function of~(\ref{GeneralOpt}).
In the pseudo code, the favorite user is identified by comparing $o_k^n$ which is defined as the objective function of~(\ref{GeneralOpt}) computed for the current channel allocation as well as the allocation of sub-channel $n$ to user $k$. After the channel allocation phase, we follow the harvesting ratio optimization phase to optimize the harvesting ratios of the users based on their allocated channels (lines 25-31). Denote $\hat{R_k}$ as the data rate of user $k$ if we change $\mu_k$ to $\hat{\mu}_k$. We also have $\hat{E}_k={T_d}(\sum\nolimits_{l \in \mathcal{K},l\ne k} {\rho _l}{\mu _l}+\rho_k\hat{\mu}_k)$ and $\hat{I}_k=\sum\nolimits_{l \in \mathcal{K},l\ne k} P_l^{tr}({\mu _l}){I_{l,m}}+ P_k^{tr}({\hat{\mu} _k}){I_{k,m}}$. As shown in the algorithm, we fist sort the users in an increasing order according to their
energy harvesting rates. Then, for each user, we locally optimize the harvesting ratio while taking into consideration of the current
$\hat{R_k}$, $\hat{E}_k$, and $\hat{I}_k$.

\textbf{Complexity Analysis:}
The optimal solution for the joint sub-channel allocation and energy harvesting optimization in the network necessitates an exhaustive search in order to find the optimal sub-channel allocation for the $K$ devices. The complexity of this exhaustive search exponentially grows as $\mathcal{O}(K^N)$. However, we can easily observe that the complexity of our proposed joint sub-channel allocation and energy harvesting optimization (INSTANT) algorithm corresponds to $\mathcal{O}(K\ast N)$. In other words, the complexity of our proposed algorithm is much lower than the exhaustive search method.
% Note that the computational complexity of the channel allocation in problem~(\ref{GeneralOpt}) is exponential and corresponds to $\mathcal{O}(2^{K\ast N})$.

\section{Simulation Results}\label{sec:simultion}
\begin{figure} [t]
	\centering
	\includegraphics[width=9cm,height=6cm]{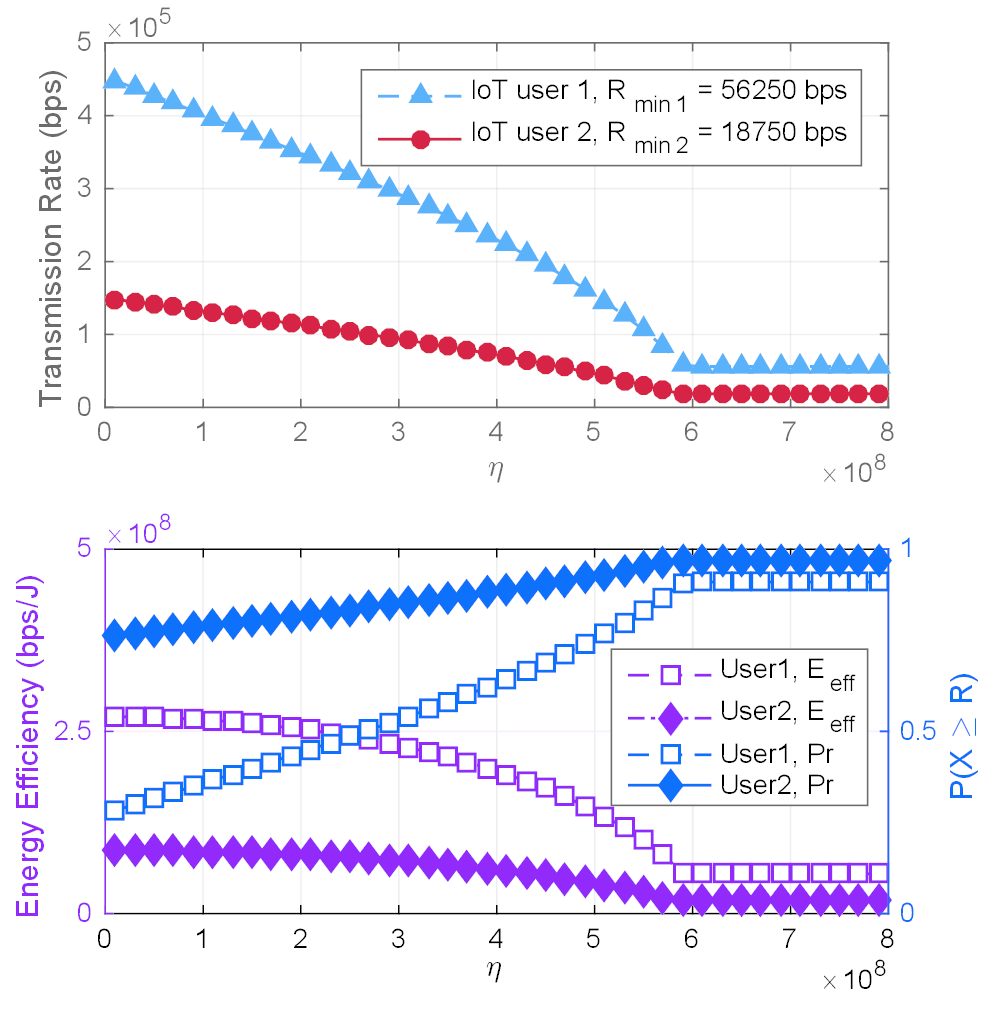}
	\caption{The transmission rates of users, and EE-SE tradeoff versus deign parameter ($\eta$). Data in buffer has a uniform distribution.}
	\label{fig:TradeOffEEandSE2in1Simulation}
\end{figure}
In this section, we evaluate the performance of the proposed optimization framework for the CR-based IoT network. We further evaluate the effectiveness of our proposed INSTANT algorithm. The OPTI toolbox~\cite{OPTI} is adopted to solve Eq.~(\ref{GeneralOpt}) by using the NOMAD\cite{NOMAD} solver, which is a global MINLP solver and uses the Mesh Adaptive Direct Search (MADS) algorithm. All simulations are executed on a computer with INTEL CPU 3.07 GHz and 10 Gb memory. The NOMAD solver and the INSTANT algorithm are run in the MATLAB 2015b on the Windows 10 operating system.
The channel gains are modeled as $h_{k,n}= Z d^{-\beta}_{k,n}$, where $Z$ is randomly generated according to the Rayleigh distribution, $d$, the geographical distance between the transmitter and receiver, is selected uniformly between 0 m to 50 m, and $\beta$, the path-loss exponent, is set to 3. Moreover, the bandwidth of each sub-channel is 62.5 kHz, the interference threshold of the licensed user is $5\times10^{-13}W$, and the noise power is $10^{-13}W$ (or $-100$ dbm) in our simulation analysis.
\begin{figure} [t]
	\centering
	\includegraphics[width=9cm,height=6cm]{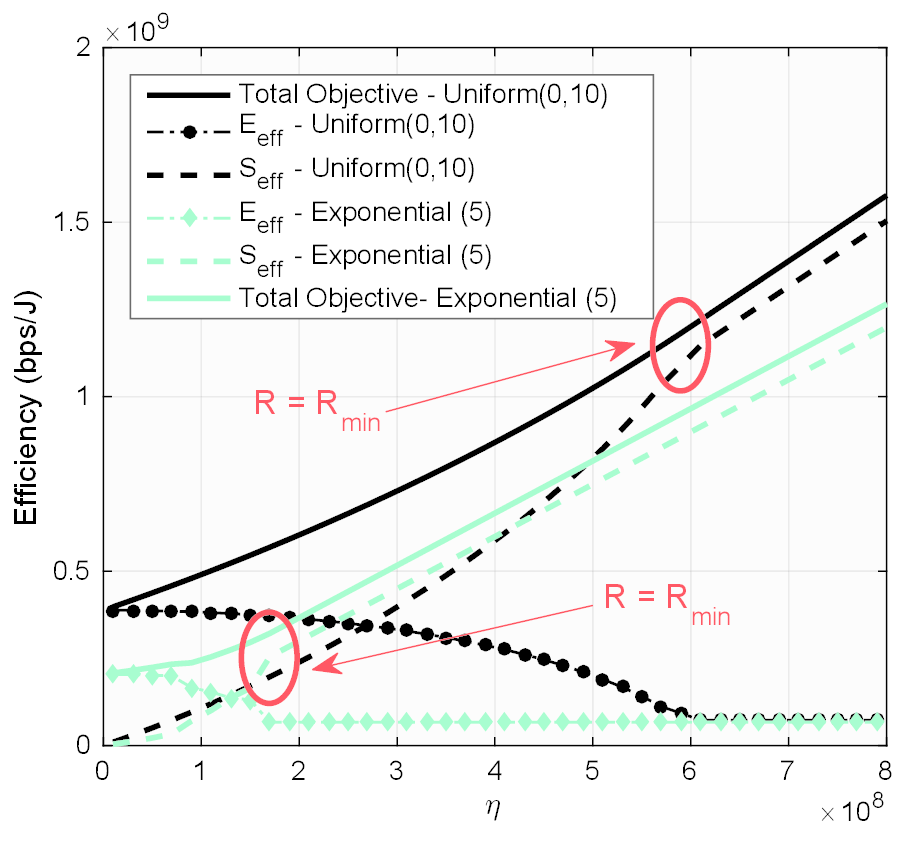}
	\caption{The optimal objective function, EE and SE versus $\eta$ for both exponential and uniform distributions of data in buffers.}
	\label{fig:ObjectivessssYY2Simulation}
\end{figure}

Fig.~\ref{fig:TradeOffEEandSE2in1Simulation} illustrates a tradeoff between EE and SE of the network. We consider two users along with four available sub-channels. The energy of sensing, residual energy and energy of idle for both users are $E^{sen} = 2 mJ$, $E^{res}= 3 mJ$ and $E^{idle}= 1 \mu J$, respectively. The data of the users in their buffers follow a uniform distribution. The first subplot clearly shows that the transmission rates for both users decrease to their minimum rate constraints as $\eta$ increases. In fact, the higher $\eta$ results in the higher weight of SE, and thus lowers the EE and transmission rates. The two users also achieve different transmission rates due to different channel conditions and harvesting rates. The second subplot of Fig.~\ref{fig:TradeOffEEandSE2in1Simulation} explicitly shows the tradeoff between EE and SE. The x-axis is the  parameter $\eta$ which is selected to be identical for both users. The left y-axis represents the energy efficiency, i.e., the first term of the objective function in Eq.~(\ref{GeneralOpt}). The right y-axis is the spectral efficiency term represented by $P(X_k \geq R_k)~\forall k$. The purple curves reflect the EE of users, where increasing $\eta$ reduces the EE to their minimum levels. The energy efficiency reduction arises from the fact that the higher $\eta$ puts the more weight on the SE term. Thus, the blue curves show that the SE of the users grow to their maximum amount by increasing $\eta$. The higher $P(X_k \geq R_k)$ implies the higher SE, but this ensures that the users do not receive rates more than their available bits in their buffers.
\begin{figure} [t]
	\centering
	\includegraphics[width=9cm,height=8cm]{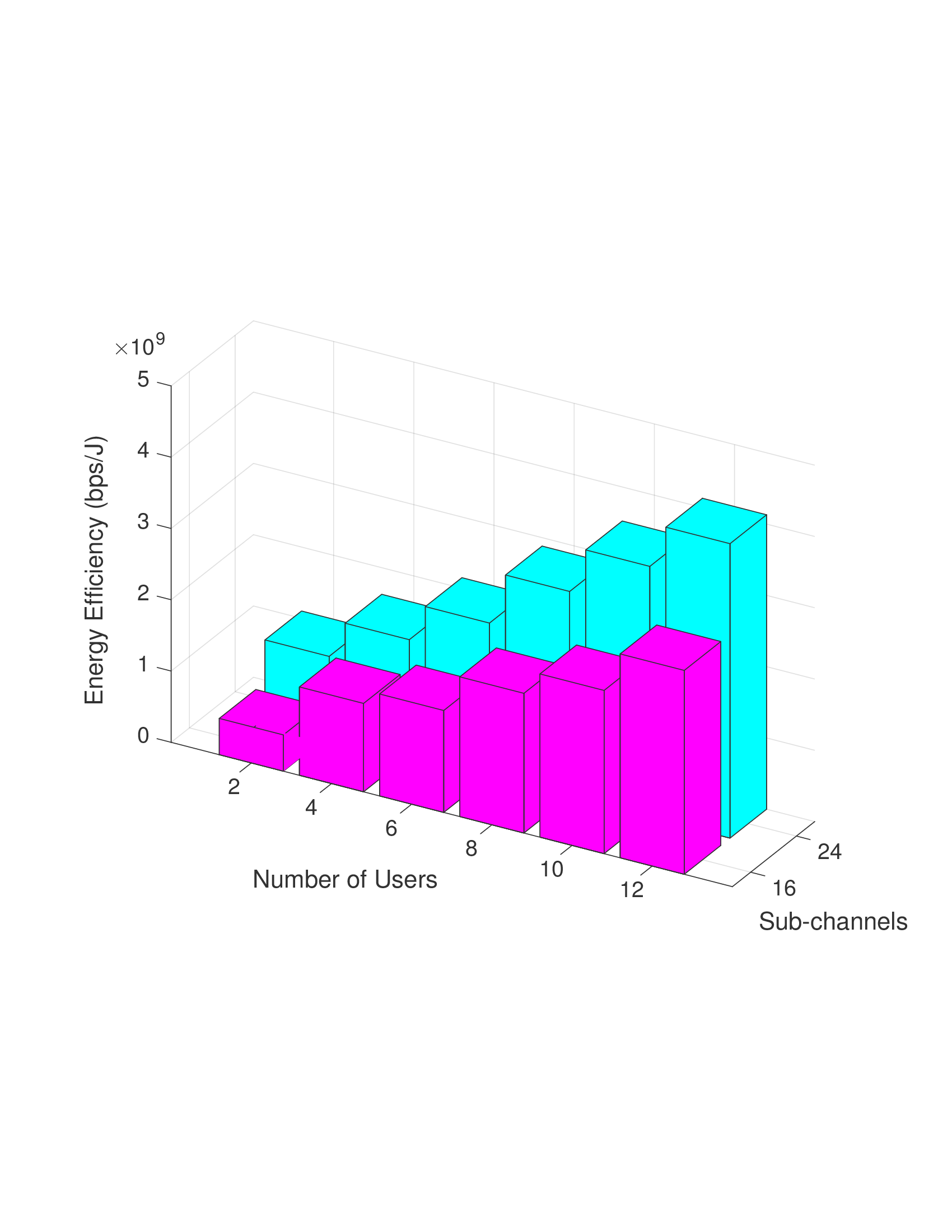}
	\caption{The energy efficiency versus the number of users and available sub-channels.}
	\label{fig:BargraphUsersAndSubchannels}
\end{figure}

Fig.~\ref{fig:ObjectivessssYY2Simulation} shows the efficiency of the network, where the number of randomly generated bits in users' buffers follow exponential and uniform distributions. The solid lines are the total objective function in Eq. (\ref{GeneralOpt}), which experiences a sharp increase by incrementing $\eta$ (identical for both users). When the weight of SE grows, the SE increases exponentially until the transmission rates of users reach their minimum thresholds. Beyond this point which is shown by red ellipse, the objective function increases with $\eta$.
% In fact, the best tradeoff for EE and SE is achieved when the EE and SE curves cross over each other. 

Fig.~\ref{fig:BargraphUsersAndSubchannels} shows the effect of the number of users and available sub-channels on the total energy efficiency. As shown in this figure, there are two scenarios in which the total numbers of available sub-channels are $N=16$ and $N=24$, respectively. For a fixed number of users, the total energy efficiency of the network grows by increasing the number of available sub-channels allocated to users. Meanwhile, the x vector represents the number of users that varies from $K=2$ to $K=12$. In fact, the higher number of users results in the higher amount of data for transmission and also more freedom for the optimizer to choose users with better channel gains. Thus, increasing the number of users lead to improving the total energy efficiency for both $N=16$ and $N=24$ scenarios.

\begin{figure} [t]
	\centering
	\includegraphics[width=9cm,height=11cm]{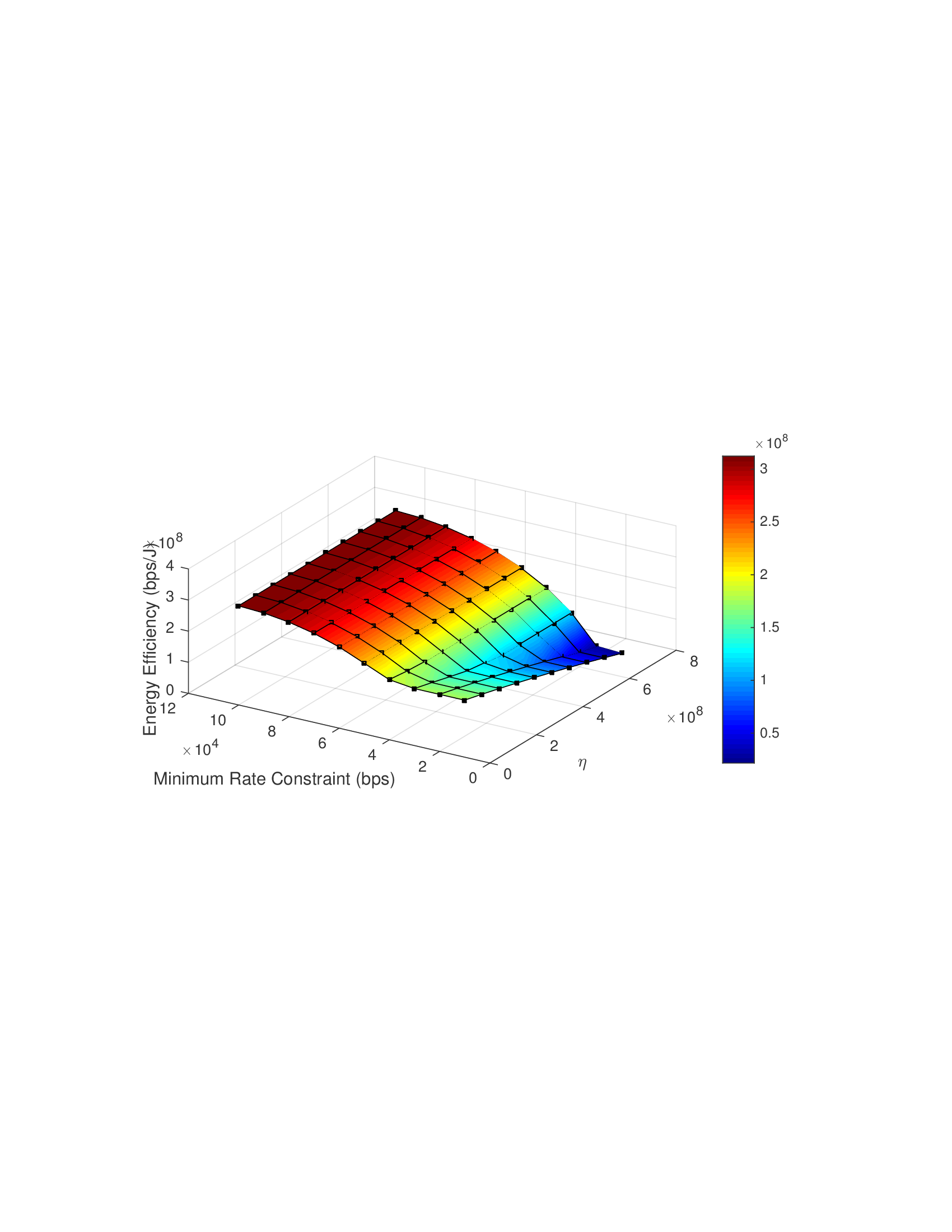}
	\caption{The energy efficiency versus minimum data rate constraint of users (identical for both users) and tradeoff parameter ($\eta$).}
	\label{fig:3DEEandRateCandEta}
\end{figure}

Fig.~\ref{fig:3DEEandRateCandEta} illustrates the total energy efficiency versus minimum rate constraint and $\eta$. 
As shown in this figure, the energy efficiency improves by increasing the minimum data rate. 
This observation is due to the fact that the transmission rate of the users must increase to satisfy $C1$ in Eq. (\ref{GeneralOpt}). In the low minimum data rate region, as $\eta$ grows, the SE term of the objective obtains a higher weight as compared to the EE term, and thus the total energy efficiency experiences a decline.
However, EE remains nearly constant for higher minimum rate constraints and does not react to the increment in values of $\eta$. This phenomenon stems from the probability term $P({X_k} \geq {{R_k}({\mu _k},{g_{k,n}})} t_k^{tr}({\mu _k}))$ in the objective function. In other words, in the high minimum data rate region, the probability $P({X_k} \geq h(E(X_{k})) t_k^{tr}({\mu _k}))$ tends to zero, and thus the SE term has no impact on the optimization problem regardless of the value of $\eta$, and it remains steady as it is shown in the figure. Note that the lowest energy efficiency occurs for the highest $\eta$ and the lowest minimum data rate constraints. 

\begin{figure} [t]
	\centering
	\includegraphics[width=9cm,height=6cm]{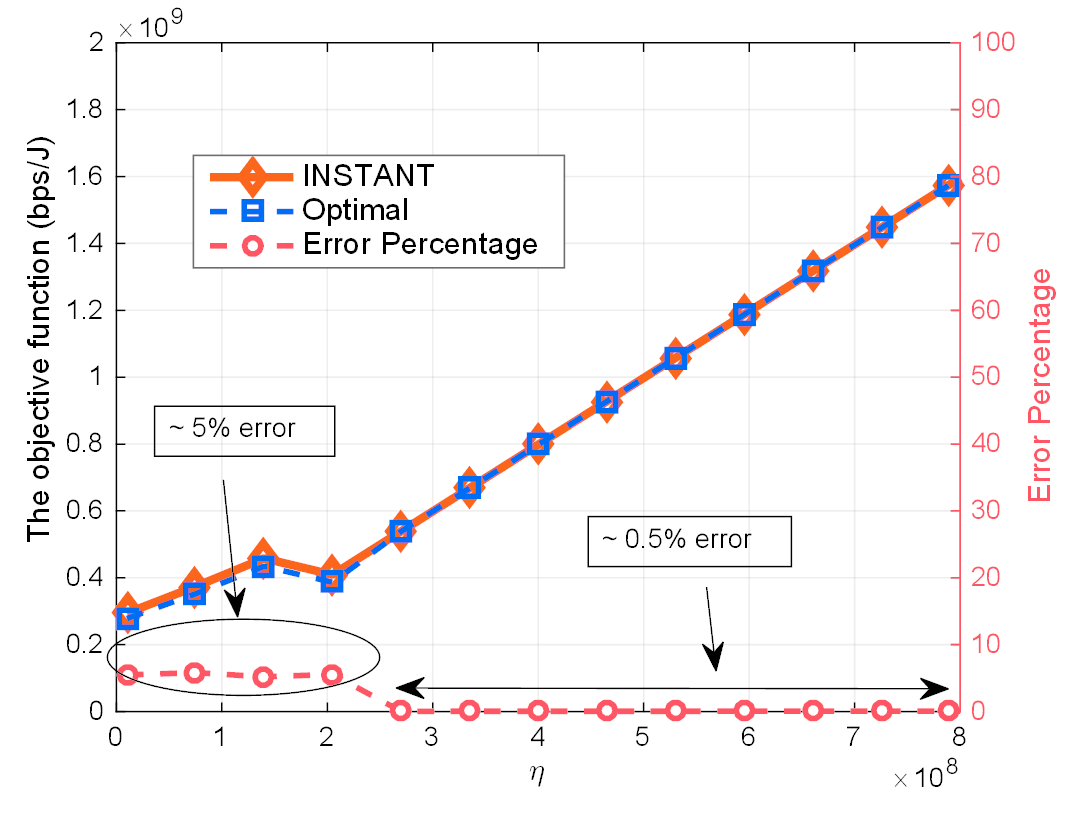}
	\caption{The objective function and error percentage versus $\eta$. The comparison between INSTANT and the optimal solution. $K=4$ and $N=8$. }
	\label{fig:INSTANTversusOptWithError}
\end{figure}

\begin{figure} [t]
	\centering
	\includegraphics[width=9cm,height=8cm]{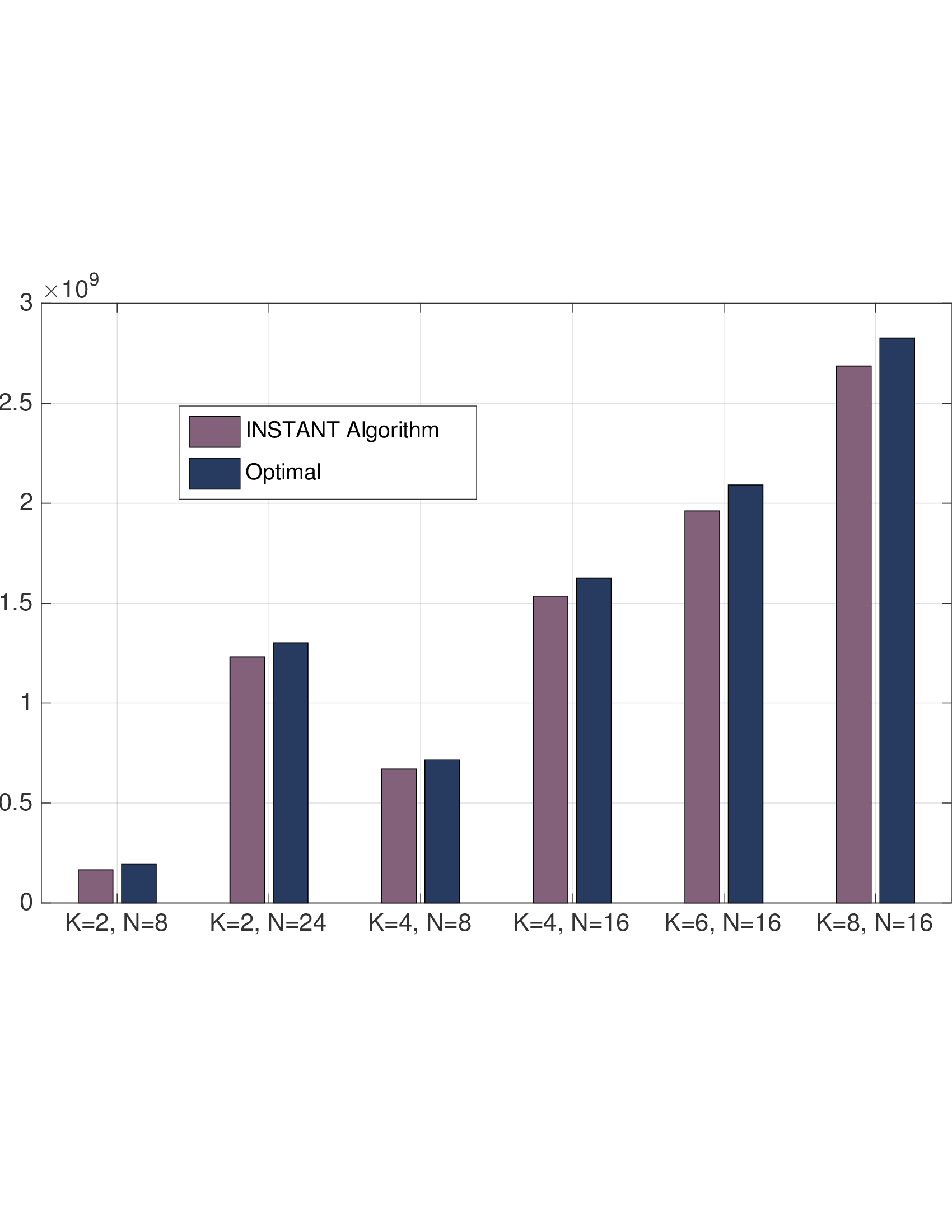}
	\caption{Energy efficiency comparisons between INSTANT and the optimal approach for different scenarios.}
	\label{fig:BargraphCompare1}
\end{figure}

%To validate the effectiveness and accuracy of the INSTANT
Fig.~\ref{fig:INSTANTversusOptWithError}, Fig.~\ref{fig:BargraphCompare1}, and Table~I are presented to evaluate the effectiveness and accuracy of the proposed INSTANT algorithm. In particular, we compare performance of INSTANT and the optimal approach in Fig.~\ref{fig:INSTANTversusOptWithError} for a network of $K=4$ users and $N=8$ available sub-channels. As shown in this figure, the objective function of Eq.~(\ref{GeneralOpt}) increases as $\eta$ grows for both INSTANT and optimal approaches. The error percentage shown on the right y-axis also presents the performance gap of INSTANT and the optimal solution for different cases. For the lower values of $\eta$ (as shown by the ellipse), INSTANT algorithm achieves the optimal result with only less than $5\%$ error. As $\eta$ increases, the results obtained by INSTANT are within $0.5\%$ of the optimal result. 
\begin{table*} \label{ComparedTable} \caption{ The computational time complexity comparisons.}
	\centering	
	\begin{tabular}{|l|l|l|l|l|l|l|}
		\hline
		& \textbf{K=2, N=8} & \textbf{K=2, N=24} & \textbf{K=4, N=8} & \textbf{K=4, N=16} & \textbf{K=6, N=16}  & \textbf{K=8, N=16}  \\ \hline \hline
		\textbf{INSTANT Algorithm} & 0.096 sec         & 0.127 sec         & 0.191 sec         & 0.304 sec & 0.445 sec & 0.612 sec          \\ \hline \hline
		\textbf{Optimal}    & 0.954 sec         & 7.804 sec         & 336.847 sec       & 990.135 sec   & 7030.431 sec   & 52728.882 sec      \\ \hline 
	\end{tabular}
\end{table*}
The comparison of computational time between the optimal approach and INSTANT for different scenarios is shown in Table~I. There are six scenarios in which the number of users and available sub-channels grows from $K=2$, $N=8$ to $K=8$, $N=24$, respectively. While INSTANT provides the sub-optimal solution within less than a second, the computational time of the optimal method grows very fast. The reason behind this arises from the exponential complexity of the optimal approach while INSTANT achieves accurate results with polynomial complexity. Moreover, Fig.~\ref{fig:BargraphCompare1} compares the energy efficiency achieved by INSTANT and the optimal method for the aforementioned scenarios; they are rather close. In particular, EEs achieved by INSTANT are $97.95\%$, $97.72\%$, $97.61\%$, $97.21\%$, $96.51\%$, and $96.16\%$ of the corresponding optimal EEs, for $\{K=2,~N=8\}$, $\{K=2,~N=24\}$, $\{K=4,~N=8\}$, $\{K=4,~N=16\}$, $\{K=6,~N=16\}$, and $\{K=8,~N=16\}$, respectively.

\section{Conclusion}\label{sec:conclusion}
In this paper, we have proposed a novel system model for CR based IoT by wireless energy harvesting and cooperative spectrum sensing to tackle two vital challenges of an IoT network, i.e., supplying adequate energy to operate the network in a self-sufficient manner, and providing enough radio spectrum for massive increase of devices. More importantly, we have formulated an MINLP problem to maximize the tradeoff between EE and SE while taking into consideration of practical limitations. Moreover, we have proposed a low complex heuristic algorithm, called INSTANT, to solve the sub-channel allocation and energy harvesting optimization problem. We have shown that INSTANT is able to obtain near optimal solution with high accuracy while having polynomial complexity.

\bibliography{EnergyEfficiencyandSpectralEfficiencyTradeoff}
\bibliographystyle{IEEEtr}

\end{document}